%% file: psat.tex
\documentstyle[11pt,epsf,subfigure]{article}

\newtheorem{definition}{Definition}
\newtheorem{theorem}{Theorem}
\newtheorem{lemma}{Lemma}
\newtheorem{corollary}{Corollary}

\newcommand{\ee}{\end{equation}} 
\newcommand{\be}{\begin{equation}} 
\newcommand{\ec}{\end{center}} 
\newcommand{\bc}{\begin{center}} 
\newcommand{\eea}{\end{eqnarray}} 
\newcommand{\bea}{\begin{eqnarray}} 
\newcommand{\bd}{\begin{description}} 
\newcommand{\ed}{\end{description}} 
\newcommand{\bi}{\begin{itemize}} 
\newcommand{\ei}{\end{itemize}} 

\begin{document}
 \bibliographystyle{apalike}
\baselineskip 16pt

\title{\bf A linear Complementarity Theorem to solve \\any  Satisfiability Problem  in conjunctive normal form \\in polynomial time}
\author{ Giacomo Patrizi \\
 Dipartimento di Scienze Statistiche \\
 Sapienza Universita' di Roma, Italy \\
 e-mail giacomo.patrizi@uniroma1.it; alternative: dove4fai@gmail.com}
\input psfig
\maketitle
\mbox{}

\abstract
{ \it
Any satisfiability problem in conjunctive normal form can be solved in polynomial time by reducing it to a 3-sat formulation and transforming this to a Linear Complementarity problem (LCP) which is then solved as a linear program (LP). Any instance in this problem class, reduced to an LCP may be solved by a  complementarity theorem,  whenever certain necessary and sufficient conditions  hold. The proof that these conditions will be  satisfied for all problems in  this class this is the contribution of this paper and this derivation requires a nonlinear Instrumentalist methodology rather than a Realist one, confirming the advantages of  a Variational Inequalities implementation.}                         
\baselineskip 16pt

\input Introduction

 \input PreliminaryResults

\input Reducing_sat

\input MainResults

\input Complexity

\input Conclusions

\input ref.tex

\end{document}

%% file: Introduction
\section{Introduction}\label{33_Introduction}

Any satisfiability problem in conjunctive normal form may be  reduced to a 3-sat formulation which can then be  expressed as a Linear Complementarity problem. Certain necessary conditions ( or axioms, if cited formally) must hold to ensure that the LCP may be solved as  a linear program ~\cite{MO79}, which can be derived by applying a simple theorem of formal matrix operations. Such a nonlinear  Instrumental methodology rather than a realist interpretation should be adopted, for the latter is  too limited. This more general methodology permits to generalize the axiom structure of the relevant class of problems and ensures that correct solutions be derived  ~\cite{DJ77} see section \ref{mainresults}.

The aim of this paper is to prove that there exists a transformation  which is always solvable for any satisfiability problem  in conjunctive  normal form, with an arbitrary number of literals and clauses, to a linear programming problem, defined over the set of rational numbers, which is bounded in the number of operations required for the transformation by a polynomial in the size of the problem.

The problem is solved as a linear program by a number of operations bounded by a polynomial defined over the finite input length of the problem, given a reasonable encoding scheme. As a linear program (LP) is a problem that can be solved in polynomial time in the length of the input  ~\cite{Khachian79}, ~\cite{Karmarkar84}, \cite{YY97}, it follows that the satisfiability problem must also be solvable in polynomial time, if the necessary conditions of the theorem ~\cite{MO79} hold for this class of problems.

Further, it is also shown that such a linear program  will always have a solution, indicating the solution to the satisfiable problem, or provides an easily recognizable solution to the linear program, proving that the  problem is falsifiable. This ensures that the polynomial reduction is well defined and in line with the concept of an algorithm as an effectively computable procedure \cite{Curry63}.
  
The polynomial solvability of the satisfiability problems is derived from two theorems and an important well founded mathematical methodology. The theorem to solve the LCP,  ~\cite{MO79}, was formulated nearly 50 years ago and it can be verified that this theorem is correct as it has been indicated that no counter examples or contradictions have ever reached the author of the original theorem  ~\cite{MO05}. The second theorem, to ensure that the conditions of the theorem be verified and the LP be solvable,  is a very simple exercise in Matrix Computations.

The following notation is used. All matrices and vectors are considered real. The transpose of a matrix $A$ or vector $ v $  are indicated by $A^T, (v^T)$.  $I$ is the identity matrix  and $e$ is a column vector of ones. Different matrices indicated with the same base letter are distinguished by superscripts.

To avoid any eventual confusion between vector and scalar notation, as is usual in Variational Inequalities derivations  all the formulations are expressed exclusively as vectors and matrices, unless the indication is evidently a digit or a scalar, in line with the handling of convex bodies and inner products, ~\cite{GM88}.

The outline of the paper is as follows.  In the next section some Preliminary results will be given to make the paper self sufficient, then in section 3 the derivation from any satisfiability problem a proper 3-sat problem and its transformation into an LCP will be proved. In section 4 the Main Results will be formulated and proved, while in section 5 the complexity of the algorithms are derived demonstrating the polynomiality of this formulation. Finally in section 6 the relevant conclusions will be drawn.

%% file: PreliminaryResults
\section{Preliminary Results} \label{a_PreliminaryResults} \label{preliminary}

The aim of this section is to present a number of definitions and results which are well known, but ensures that the terms used in the derivations are consistent.

Consider a satisfiability problem in conjunctive  normal form, ~\cite{SvW88}
\begin{definition} 
\label{def2.5}  Let U be a set of symbols over an alphabet, called propositional variables, and denoted by $U = \left \{ u_i : i = 1, 2, ..., n \right \}$. Let F be an expression from the language $L\in \Sigma^*$ called a propositional formula, built from logical connectives defined by the rules of the propositional calculus. An assignment of truth values to the propositional variables occurring in a propositional formula F is a function $v:U\rightarrow \{0,1\}$. The truth value of the expression is determined by the rules of the propositional calculus and is denoted ( with a slight abuse of notation) by v(F). A propositional formula F is satisfiable if there is at least one assignment such that v(F) = 1 and otherwise it is falsifiable. 
\end{definition}

\begin{definition} 
\label{def2.6} A literal $l_j, j = 1, 2, ..., n$, is either a propositional variable or the negation of a propositional variable. A propositional formula F is in conjunctive normal form, if it is in the format $F = c_1 \wedge c_2 \wedge \ldots \wedge c_m$  for some $m\geq 1$, where $\wedge $ is the intersection operator. The $ c_i $, $ i= 1,2,...,m $ are called clauses and are in the format  $l_1\vee l_2 \vee \ldots \vee l_n$ , where $ \vee $ is the union operator and each $l_j, \quad j=1,2,...,n $  is a literal.
\end{definition}

\begin{definition} 
\label{def2.7} A  satisfiability problem  is a propositional formula in conjunctive normal form and is an instance of the language considered.
\end{definition}

\begin{definition}
\label{def2.8} A  3-sat problem  is a satisfiability problem reduced from a satisfiability propositional formula in conjunctive normal form with clauses with at most 3 literals, which are not repeated within a clause and is an instance of the language considered \cite{GM79}.
\end{definition}

\begin{definition}
\label{def2.2}
Let  $ M \in {\bf R}^{n \times n}$ be a square matrix and  $q \in {\bf R}^n $ be an affine vector. Also let  $u  $ be an $n$ dimensional non negative vector to be determined. The LCP may be stated:
\begin{equation}
\label{eqn2.5}
Mu + q \geq 0, \quad u \geq 0,\quad u^T(Mu+q) = 0
\end{equation}
\end{definition}

 Let $Z^1, Z^2 $ be square  $Z$-matrix of the same dimension  $ n \times n $  with non positive off-diagonal elements \cite{FiedlerPtak62},  $ r, s, c  $ be nonnegative vectors of order $n$. The vector $ u $ may be determined by solving a suitable Linear Program [LP].

\begin{theorem} (\cite{MO79}, theorem 3)
\label{theorem3.1}\\
The LCP (\ref{eqn2.5}) has a solution if and only if the LP:
\begin{eqnarray}  
              & Min & (r^T + s^TM)u            \label{eqn3.4a} \\
              & s.t.&  Mu + q \geq  0,          \label{eqn3.4b} \\    
              &     &  u \geq  0                \label{eqn3.4c}
 \end{eqnarray}                                                          
\noindent is solvable for some $ r, s  \in  {\bf R}^{n}$ which must satisfy the following conditions:
\begin{eqnarray}
        &(a)& MZ^1 = Z^2 + qc^T                 \label{eqn3.5a}\\
        &(b)& r^TZ^1 + s^TZ^2 \geq  0           \label{eqn3.5b}\\
        &(c)& r^TZ^1 + s^TZ^2 + c^T > 0         \label{eqn3.5c}\\
        &(d)& r + s  > 0                        \label{eqn3.5d}\\
        &   & c, r, s \geq 0  \quad  Z^1, Z^2 \in Z
\end{eqnarray}
for some vector $ c \in {\bf R}^n $  and some matrices $  Z^1, Z^2 \in{\bf R}^{n \times n} $. Furthermore, each solution of the LP solves the LCP.
\end{theorem}

If  the transformation is well defined then the resulting LP has an optimal solution so the complementary slackness condition between the primary variables and the dual slack variables extends to the primal slacks  ~\cite{PG91}. This is just the condition required to solve the LCP. Therefore  the proposed algorithm is  an effectively computable procedure ~\cite{Curry63}.

%% file: Reducing_sat
\section{Reducing satisfiability to 3sat and to LCP}  \label{red_lcp}

A satisfiability problem given in definition \ref{def2.7}, containing $n$ literals and $m$ clauses, without any limitation as to the number of literals present in a clause  can be transformed to a 3-Satisfiability problem as in definition \ref{def2.8} in polynomial time ~\cite{GM79}.

The upper bound to the total number of clauses in the 3-satisfiability formulation  will be less than or equal to $nm$  ~\cite{GM79}. Further the number of added literals is bounded by $nm$ since an auxiliary literal is required for each new clause. Hence the size of the 3-satisfiability problem can be considered as composed of $ N \leq n+nm $ literals and $ M \leq nm $ clauses.

The transformed problem can also be formulated as a system of linear inequalities over a set of boolean variables. Denote the set of propositional variables of the satisfiability problem by an equivalent set of boolean variables, $X$  which take on values \{0,1\}. Then to formulate a literal write the corresponding boolean  variable as $x_i$ if the literal of the propositional variable $u_i$ is in affirmative form  and by $(1 - x_i)$ if it is in the negation form. Then every clause forms an inequality, given as a sum of terms $x_i$  or $(1 - x_i)$ depending on the literals.   

Assign the value of $x_i = 1$  if $u_i$  is assigned a truth value of "true" and $x_i =  0$, otherwise. A clause is true if the sum of terms in the boolean expression formed from the literals has value greater than, or equal to one and the clause is false otherwise. 

By construction any clause is represented by a sum of three terms composed by $x_i$ or $(1 - x_i)$ for $i = 1,2,...,N$, where $N$ is the number of literals present in the propositional formula of the transformed problem. 

In a clause, let the index set of those literals expressed as affirmative propositional variables be given by $I$ while that containing negated literals be indicated by $J$. As we are dealing exclusively with satisfiability problems with three literals in each clause. Any clause of such a satisfiability problem can be represented as an inequality:

\begin{equation}
\label{eqn3.5}
\sum_{i\in I}x_i + \sum_{j\in J}{\left (1-x_j \right )}\geq 1                                  
\end{equation}
or by collecting terms, the inequality can be written as:

\begin{equation}
\label{eqn2.6}
              a^Tx + b \geq 0                                                                  
\end{equation}
where $a$ is a vector of elements $\left \{a_k\right \} = -1, 0,+1,  k = 1, 2, ... , N$,  and $b = \#(J) -1 \quad $ \cite{Jeroslow89}.

In a propositional formula expressed in  conjunctive  normal form, all clauses must be simultaneously true for the formula to be true. It follows that expressing each clause as an inequality (\ref{eqn2.6}), the system of inequalities will be feasible whenever a consistent evaluation of the boolean variables is given which corresponds to a satisfiable assignment of the propositional variables. 

 An equivalent representation  may be expressed, without loss of generality, as a system of inequalities:

\begin{equation}
\label{eqn2.7}
Cx + b \geq 0
\end{equation} 
where the vector $ x $ has $x_i = \{0,1\} $ and integer, $ \forall i = 1, 2,..., N $, the elements of the vector b, also integer, with  $-1\leq b_m \leq 2,  \forall m \in  M $ and the matrix C is an $M\times N$ dimensional matrix with only three elements different from zero  per row, whose values are: -1, or +1.

This may be easily represented in the form of an LCP of dimension $ (N+M) \times (N+M)$, by introducing a vector of artificial nonnegative variables $ \Gamma \in R^M $
 
\begin{eqnarray}
\left (\begin{array}{cc}
        -I & 0\\
         C & 0
        \end{array}
\right )
\left (
        \begin{array}{c}
        x\\
        \Gamma
        \end{array}
\right ) +
\left (
        \begin{array}{c}
        e\\
        b
        \end{array}
\right ) & \geq & 0     \label{eqn2.8}\\
\left (
        \begin{array}{c}
        x\\
        \Gamma
        \end{array}
\right ) & \geq & 0     \label{eqn2.9}\\
\left (
        \begin{array}{cc}
        x^T, \Gamma^T
        \end{array}
\right )^T
\left ( \left (
                \begin{array}{cc}
                -I & 0\\
                 C & 0
                \end{array}
        \right )
        \left (
                \begin{array}{c}
                x\\
                \Gamma
                \end{array}
        \right )+
        \left (
                \begin{array}{c}
                e\\
                b
                \end{array}
        \right )
\right ) & = & 0        \label{eqn2.10}
\end{eqnarray}                                                                                                               
\begin{lemma} 
\label{lemma2.1} The 3-sat problem is satisfiable if and only if the LCP (\ref{eqn2.8}), (\ref{eqn2.9}), (\ref{eqn2.10}) has a solution.
\end{lemma}

{\noindent \sc Proof}: ($\Rightarrow$) Suppose that the 3-sat problem is satisfiable and indicate the truth values of the propositional variable $u_i \in U$ by  $u^*$ where $u_i^*$  may assume values true or false. In the LCP (\ref{eqn2.8}), (\ref{eqn2.9}), (\ref{eqn2.10})  take $x_i=1$ if $u_i^*$  assumes the value of true and $x_i=0$ otherwise. 

 Such a solution is a feasible solution to the set of inequalities (\ref{eqn2.8}), since $x\in \{0,1\}^N$ and integer. The solution is also complementary, as it satisfies (\ref{eqn2.10}), since in the first $N$ inequalities either $x_i = 0$ or  $x_i = 1, i=1,...,N $ and for these the inequality yields $x_i (-x_i + 1) = 0$. Thus the complementarity condition holds and the solution is a solution to the LCP on imposing $ \Gamma = 0$.

($\Leftarrow$) Let the LCP have a solution, $(x' ,\Gamma')$. It must be a feasible solution to the inequalities (\ref{eqn2.8}), so $Cx'+ b \geq 0$ and $0 \leq x_i'\leq  1, \forall \, i = 1, 2,..., N$. By the complementarity condition each element of the inner product between the two vectors must be null, as both vectors are nonnegative, so $x_i'(-x_i' + 1) = 0$, $ i = 1, 2,..., N$, for the solution to be complementary and to satisfy (\ref{eqn2.10}). Thus, $x_i' =  0$ or $x_i' = 1$. Whatever value of the vector $\Gamma'$, there is no loss in generality to take $ \Gamma' = 0 $ as this will not affect the solution of the LCP. By applying the assignment $u_i^*$  is 'true' to  the propositional variable,  if $x_i = 1, i= 1, 2, ..., N $, and 'false' otherwise, each clause is satisfiable, since each will result to have a truth value 'true', by construction as (\ref{eqn2.7}) is feasible.  Therefore the satisfiability problem is satisfiable. $\Box$
                 
Thus, without loss of generality, any satisfiability problem can be transformed into an LCP, which has some additional structure which will be useful below in establishing the required results.  

 To be an interesting problem, the LCP (\ref{eqn2.8}), (\ref{eqn2.9}), (\ref{eqn2.10}) must have at least one element of the vector $b$ negative, for if  $b\geq 0$, the satisfiability problem is trivially satisfiable with all the propositional variables set to 'false'.  The vector $b\geq  0$ in (\ref{eqn2.7}) occurs only when all the clauses in the propositional formula in conjunctive normal form contain a literal which is a negated propositional variable. In this case, the null solution is an obvious solution to the satisfiability problem. Thus a satisfiability problem is not trivial if there is at least one clause with all affirmative literals, so that, for this clause on transforming it to our notation, there results: $b_i =-1$.

 Similarly, there must be at least one clause  with all its  literals negated. Under the transformation (\ref{eqn2.6}) this means that an inequality must have all its non null coefficients negative, as otherwise, $x = e$ will be a solution to the set of inequalities. This would correspond to the case where every clause contains one or more literals in the affirmative  form, so that assigning to all the propositional variables the value 'true' will render the problem satisfiable.

\begin{definition}
\label{def2.8new} The following sets are to be distinguished:

\begin{itemize}

\item The set of clauses in which all coefficients of the propositional variables are non negative,\\
    $ K = \left \{k\in M |  c_{kj} \neq 0 \Rightarrow c_{kj} = 1, \quad \forall j = 1, 2, ... , N, \quad b_k = -1\right \},$

\item The set of clauses in which all the coefficients of the propositional variables are non positive,\\ 
   $ L = \left \{l\in M | c_{lj} \neq 0 \Rightarrow c_{lj} = -1, \quad \forall j = 1, 2, ... , N, \quad b_l= 2 \right \},$

\item The set of clauses in which all the coeffficients of the propositional variables, but one  are non positive, \\
             $  Q = \left \{q\in M |\quad \exists \, c_{qi} = 1  \quad and \quad c_{qj} \neq 0 \Rightarrow c_{qj} = -1, \quad \forall i\neq j = 1, 2, ... , N, \quad b_q = 1 \right \},$

\item The set of remaining clauses in which all the coefficients of the propositional variables but one  are non negative:  \\
$  R = \left \{ r \in M |\quad \exists \, c_{ri} = - 1 \quad and \quad c_{rj} \neq 0 \Rightarrow c_{rj} = 1, \forall i\neq j = 1, 2, ... , N, \quad b_r = 0 \right \},$
\end{itemize}
\end{definition}

Group the inequalities by type and extend the coefficient matrix with null values and insert suitable identity matrices to form an  $ (N + M) \times (N +M) $ matrix, so  define a vector $ y \in {\bf R}^M_+ $ of artificial variables which will be essential in determining the optimal solution to the LP. Consider in (\ref{eqn2.7}) the set $M = \{K, L, Q, R\}$, where : $K\neq \emptyset,\quad   L\neq \emptyset$, since it is assumed that the satisfiability problem is not trivial. Thus the LCP (\ref{eqn2.8}) -  (\ref{eqn2.10}) can be written:

\begin{equation}
\left  (  
  \begin{array}{ccccc}
   -I_{NN} &    0  &   0   &   0   & 0\\
    C_{KN} & I_{KK}&   0   &   0   & 0\\
    C_{LN} &    0  & I_{LL}&   0   & 0\\
    C_{QN} &    0  &   0   & I_{QQ}& 0\\
    C_{RN} &    0  &   0   &   0   & I_{RR} 
  \end{array}
\right )
\left (
  \begin{array}{c}
    x_N\\
    y_K\\
    y_L\\
    y_Q\\
    y_R
   \end{array}
\right ) +
\left (
  \begin{array}{c}
    e_N\\
    -e_K\\
    2e_L\\
    e_Q\\
    0
  \end{array}
\right )            \geq  0   \label{eqn2.10bis} 
\end{equation}
\smallskip
\begin{equation}
\left  ( 
  \begin{array}{c}
    x_N\\
    y_K\\
    y_L\\
    y_Q\\
    y_R
   \end{array}
\right )             \geq  0  \label{eqn2.11}\\
\end{equation}
\smallskip
\begin{equation}
\left ( 
  \begin{array}{ccccc}
    x_N^T, & y_K^T, & y_L^T, & y_Q^T, & y_R^T
  \end{array}
\right )
\left (
  \left (
    \begin{array}{ccccc}
   -I_{NN} &    0  &   0   &   0   & 0\\
    C_{KN} & I_{KK}&   0   &   0   & 0\\
    C_{LN} &    0  & I_{LL}&   0   & 0\\
    C_{QN} &    0  &   0   & I_{QQ}& 0\\
    C_{RN} &    0  &   0   &   0   & I_{RR}
  \end{array}
  \right )
  \left (
    \begin{array}{c}
      x_N\\
      y_K\\
      y_L\\
      y_Q\\
      y_R
    \end{array}
  \right ) +
  \left (
    \begin{array}{c}
      e_N\\
      -e_K\\
      2e_L\\
      e_Q\\
      0
    \end{array}
  \right )
\right )    = 0    \label{eqn2.12}
\end{equation}
\smallskip

 Any satisfiability problem in conjunctive normal form can be represented in this fashion. For each instance, the number of elements in the sets $N$, $K$, $L$, $Q$, and $R$ may vary, but $K$ and $L$  must be  not empty for the problem to be not trivial.

%% file: MainResults
\section{Main Results}      \label{mainresults}

Any satisfiability problem can be reduced to a LCP (\ref{eqn2.10bis})-(\ref{eqn2.12}) and some additional useful matrices can be derived from the coefficient matrix of the LCP.

A nonlinear instrumental methodology must be applied to ensure that the required axioms  be valid and  applicable and therefore  permit the construction of formulable derivations  as stated in theorem \ref{theorem3.1} to build realistically solvable algorithms which will be proven in theorem \ref{theorem4.3} ensuring that the axioms are correct ~\cite{DJ77} ~\cite{BN84}.

 To satisfy the conditions of theorem \ref{theorem3.1} suitable matrices $Z^1  $ and $Z^2$  and vectors $ r, s $ must be defined for the problem and be correctly structured . As it will be evident in the  formulation and the proof of theorem \ref{theorem4.3} each column must sum conformably to a posive value, so two instrumental additional artificial rows must be added to $ M $. To achieve this, form  two row vectors $ v^T, w^T $  of order $( N + M+2) $ and two additional columns must be inserted to  render the LCP of order $ ( N + M +2) $.

To construct the vector $ v^T $ determine:

\begin{equation}
\label{eqn4.1}
 g^T_N =  e^T_{M} \left (
        \begin{array}{c}
       C_{KN}   \\
       C_{LN}   \\
       C_{QN}   \\
       C_{RN}   \\
           \end{array}
 \right )          \label{eqn4.19} 
\end{equation}

and set  $ (v^T_N)_i = -(g^T_N)_i > 0  \quad \mbox{if} \quad (g^T_N)_i < 0, \quad \forall i = 1, 2, \cdots, N  $.  The  row vector $ v^T_{N} $ is then extended to cover with all null values the columns of the matrix  $ M $ which now are  $ ( N + M + 2 ) $, rows and columns, so that the vector becomes a nonnegative row vector of order $(N + M + 2) $  and the $ (N + M + 1) $-th element is given a value of $ 1 $.

A more complex construction is necessary to  build the additional vector $ w^T $.

Consider a new matrix $ P $ of order $ M \times N$ defined from the original matrix $ M $  by selecting appropriate nonpositive  terms by the following sets

\begin{definition}
\label{def4.8} The following sets are to be distinguished:

\begin{itemize}
\item The set of clauses in which all coefficients of the propositional variables are non negative,\\
 $ K = \left \{k\in M | c_{kj} > 0 \Rightarrow c_{kj} = 0, \quad \forall j = 1, 2, ... , N \right \},$
\item The set of clauses in which all the coefficients of the propositional variables are non positive,\\
 $ L = \left \{l\in M | c_{lj} < 0 \Rightarrow c_{lj} = -1, \quad \forall j = 1, 2, ... , N \right \},$
\item The set of clauses in which all the coeffficients of the propositional variables, but one  are non positive, \\
 $ Q = \left \{q\in M | c_{qj} < 0 \Rightarrow c_{qj} = -1, \quad \forall j = 1, 2, ... , N \right \},$
\item The set of remaining clauses in which all the coefficients of the propositional variables but one are non negative: \\
 $ R = \left \{ r \in M | c_{rj} < 0  \Rightarrow c_{rj} = -1, \forall    j = 1, 2, ... , N, \right \},$
\end{itemize}
\end{definition}

So the required matrices $ Z^1 $, $ Z^2 $ can be defined as  the matrix $ P $ is nonpositive and the row vector $ w^T $ will be specified below.

Thus:
\begin{equation}
\label{eqn3.6}
M = \left (
         \begin{array}{ccccccc}
      -I_{NN}       &    0    &       0      &      0      & 0       & 0  & 0 \\
       C_{KN}       &  I_{KK} &       0      &      0      & 0       & 0  & 0  \\
       C_{LN}       &    0    &      I_{LL}  &      0      & 0       & 0  & 0  \\
       C_{QN}       &    0    &      0       &      I_{QQ} & 0       & 0  & 0  \\
       C_{RN}       &   0     &      0       &      0      & I_{RR}  & 0  & 0   \\
        v^T_{N}     &   0     &      0       &      0      & 0       & 1  & 0   \\
        w^T_{N}     &   0     &      0       &      0      & 0       & 0  & 1  
           \end{array}
  \right ),  \left (
                \begin{array}{c}
                        e_N\\
                        -e_K\\
                        2e_L\\
                        e_Q\\
                        0  \\
                        0  \\
                \end{array}
                \right )               
\end{equation}

\bigskip
\begin{equation}
 \label{eqn3.7}
Z^1 = \left (
        \begin{array}{ccccccc}
        -I_{NN}  &     0      &      0      &      0      & 0      & 0 & 0\\
         P_{KN } &     I_{KK} &      0      &      0      & 0      & 0 & 0 \\
         P_{LN}  &     0      &      I_{LL} &      0      & 0      & 0 & 0\\
         P_{QN}  &     0      &      0      &      I_{QQ} & 0      & 0 & 0 \\
         P_{RN}  &     0      &      0      &      0      & I_{RR} & 0 & 0  \\
         0^T_{N} &     0      &      0      &      0      &  0     & 1 & 0  \\
         0^T_{N} &     0      &      0      &      0      &  0     & 0 & 1  
        \end{array}
  \right ),                             
 \end{equation}

resulting in the following matrix:

\bigskip
\begin{equation}
 \label{eqn3.8}
Z^2 = MZ^1= \left (
        \begin{array}{ccccccc}
             +I_{NN}    &    0       &      0      &      0      & 0        & 0 & 0 \\
      -C_{KN} + P_{KN}  &    I_{KK}  &      0      &      0      & 0        & 0 & 0 \\
      -C_{LN} + P_{LN}  &    0       &      I_{LL} &      0      & 0        & 0 & 0  \\
      -C_{QN} + P_{QN}  &    0       &      0      &      I_{QQ}  & 0        & 0 & 0  \\
      -C_{RN} + P_{RN}  &    0       &      0      &      0      & I_{RR}   & 0  & 0 \\
           -v^T_{N}     &    0       &      0      &      0      & 0        & 1  & 0 \\
           -w^T_{N}     &    0       &      0      &      0      & 0        & 0  & 1
        \end{array}
    \right )
\end{equation}

To determine the vector $ w^T_{N} $ consider the sum of the $ N $ columns  and last $M+1 $ rows with changed sign, so the sum will be positive.

\begin{equation}
\label{eqn4.51}
 h^T_N =  e^T_{M} \left (
        \begin{array}{c}
       C_{KN} - P_{KN}   \\
      C_{LN} - P_{LN}   \\
      C_{QN} - P_{QN}   \\
      C_{RN} - P_{RN}   \\
           v^T_{N}       \\
           \end{array}
 \right ) \geq 0
\end{equation}

By construction $ h^T_N -(g^T_N + v^T_{N}) = -e^TP \geq 0 $, so the row vector $ w^T_{N} $ can then be extended to cover all the columns with null values of the row vector  rendering a nonnegative row  vector of order $ (N + M + 2) $.  Place $ 1 $ in the position $ (N + M + 2) $.
  
The  complete  the structure of the coefficient matrix $ M $  and the matrices $ Z^1$, $ Z^2$  define the instrumental LCP to derive solutions to the satisfiability problem and satisfy the conditions of theorem \ref{theorem3.1} and is given by:

\begin{equation}
 M =  \left (
          \begin{array}{ccccccc}
        -I    &    0 &       0      &      0      & 0 & 0 & 0 \\
       C_{KN} &    I &       0      &      0      & 0 & 0 & 0 \\
       C_{LN} &    0  &      I      &      0      & 0 & 0 & 0  \\
       C_{QN} &    0  &      0      &      I      & 0 & 0 & 0 \\
       C_{RN} &    0  &      0      &      0      & I & 0 & 0 \\
       v^T_{N}&    0  &      0      &      0      & 0 & 1 & 0  \\
       w^T_{N}&    0  &      0      &      0      & 0 & 0 & 1   
           \end{array}
    \right )  
    \left (
           \begin{array}{c}
             x_N\\
             y_K\\
             y_L\\
             y_Q\\
             y_R \\
             y_v \\ 
             y_w
           \end{array}
    \right ) +
    \left (
           \begin{array}{c}
             e_N\\
             -e_K\\
            2e_L\\
             e_Q\\
             0  \\
             0  \\
             0
           \end{array}
    \right )   \geq 0 \label{eqn4.123}                 
\end{equation}
\begin{equation}
\left (
           \begin{array}{c}
             x_N\\
             y_K\\
             y_L\\
             y_Q\\
             y_R \\
             y_v \\
             y_w
           \end{array}
    \right )                   \geq 0 \label{eqn4.124}
\end{equation}
\begin{equation}
 \left (
           \begin{array}{ccccccc}
             x_N^T, & y_K^T, & y_L^T, & y_Q^T, & y_R^T, y_v,y_w
           \end{array}
     \right )
           \left (
           \left (
            \begin{array}{ccccccc}
        -I    &    0 &       0      &      0      & 0 & 0 & 0\\
       C_{KN} &    I &       0      &      0      & 0 & 0 & 0 \\
       C_{LN} &    0  &      I      &      0      & 0 & 0 & 0  \\
       C_{QN} &    0  &      0      &      I      & 0 & 0 & 0  \\
       C_{RN} &    0  &      0      &      0      & I & 0 & 0  \\
       v^T_{N} &   0  &      0      &      0      & 0 & 1 & 0  \\  
       w^T_{N} &   0  &      0      &      0      & 0 & 0 & 1    
           \end{array}
           \right )
           \left (
             \begin{array}{c}
               x_N\\
               y_K\\
               y_L\\
               y_Q\\
               y_R \\
               y_v \\
               y_w
             \end{array}
           \right ) +
           \left (
             \begin{array}{c}
               e_N\\
               -e_K\\
               2e_L\\
               e_Q\\
               0  \\
               0 \\
               0
             \end{array}
         \right )           
    \right )  = 0  \label{eqn4.125}  
           \end{equation}

\bigskip
\begin{equation} 
Z^1 = \left ( 
        \begin{array}{ccccccc}
        -I_{NN}  &     0      &      0      &      0      & 0      & 0 & 0 \\
         P_{KN } &     I_{KK} &      0      &      0      & 0      & 0 & 0 \\
         P_{LN}  &     0      &      I_{LL} &      0      & 0      & 0 & 0\\
         P_{QN}  &     0      &      0      &      I_{QQ} & 0      & 0 & 0\\
         P_{RN}  &     0      &      0      &      0      & I_{RR} & 0 & 0 \\
         0^T_{N} &     0      &      0      &      0      &  0     & 1 & 0 \\
         0^T_{N} &     0      &      0      &      0      &  0     & 0 & 1    
        \end{array}
  \right ),                                 \label{eqn4.7}\\          
 \end{equation}

resulting once more in the extended form in the following matrix:

\bigskip
\begin{equation}
 \label{eqn4.8}
Z^2 = MZ^1= \left ( 
        \begin{array}{ccccccc}
             +I_{NN}    &    0       &      0      &      0      & 0        & 0 & 0 \\
      -C_{KN} + P_{KN}  &    I_{KK}  &      0      &      0      & 0        & 0 & 0 \\
      -C_{LN} + P_{LN}  &    0       &      I_{LL} &      0      & 0        & 0 & 0 \\
      -C_{QN} + P_{QN}  &    0       &      0      &      I_{QQ}  & 0        & 0 & 0  \\
      -C_{RN} + P_{RN}  &    0       &      0      &      0      & I_{RR}   & 0 & 0 \\
           -v^T_{N}     &    0       &      0      &      0      & 0        & 1 & 0 \\ 
           -w^T_{N}     &    0       &      0      &      0      & 0        & 0 & 1           
        \end{array}
    \right )                                      
\end{equation}

In (\ref{eqn4.123}) the first matrix is $ M $ and the second is the affine vector of the problem.  Then $ Z^1 $ is indicated  by  (\ref{eqn4.7}) and the product matrix $ Z^2$  is given by  (\ref{eqn4.8}).

 The  matrices   $ Z^1 $  and $ Z^2 $  indicated satisfy the conditions (\ref{eqn3.5a}) and  additional conditions can be enforced by selecting the following values:

\begin{eqnarray}
s^T_{N}  &=&  ( e^T_{N} +g^T_{N} -e^TP +  v^T_{N}  )     \label{eqn3.14}\\
s^T_K   &=&  e^TI_{KK}    \label{eqn3.14d}\\
s^T_L   &=&   e^TI_{LL}   \label{eqn3.14e}\\
s^T_Q   &=&  e^TI_{QQ} \label{eqn3.14f}\\
s^T_R   &=&  e^TI_{RR}    \label{eqn3.14g}\\
v   &=&  1    \label{eqn3.14h}\\
w    &=&  1    \label{eqn3.14p}
\end{eqnarray}

Take  $ r^T = 0 $ and $ s^T > 0 $, as indicated, so $  r + s > 0 $ then $ r^TZ^1 + s^TZ^2  > 0  $. Thus all the conditions of theorem \ref{theorem3.1} are satisfied.

 The LP  to be solved is indicated as:
               
\begin{equation}
 \mbox{Min} W  =  \left (
\begin{array} {ccccccc}
(s^T_{N} & s^T_{K} & s^T_{L} & s^T_{Q} & s^T_{R} &  1 & 1 ) 
\end{array}
 \right )
  \left (
          \begin{array}{ccccccc}
        -I_{NN} &    0   &   0      &      0      & 0      & 0 & 0 \\
       C_{KN}   & I_{KK} &   0      &      0      & 0      & 0 & 0 \\
       C_{LN}   &    0   &   I_{LL} &      0      & 0      & 0 & 0  \\
       C_{QN}   &    0   &   0      &      I_{QQ} & 0      & 0 & 0 \\
       C_{RN}   &    0   &   0      &      0      & I_{RR} & 0 & 0 \\
       v^T_{N}  &    0   &   0      &      0      & 0      & 1 & 0  \\
       w^T_{N}  &    0   &   0      &      0      & 0      & 0 & 1   
           \end{array}
    \right )
     \left (
           \begin{array}{c}
             x_N\\
             y_K\\
             y_L\\
             y_Q\\
             y_R \\
             y_v \\
             y_w
           \end{array}
        \right )       \label{eqn4.13} \\
      \end{equation}

subject to:

\begin{equation}
\left (
           \begin{array}{ccccccc}
          -I_{NN}    &    0    &       0      &      0      & 0        & 0   & 0\\
       C_{KN}       &  I_{KK} &       0      &      0      & 0       & 0    & 0 \\
       C_{LN}       &    0    &      I_{LL}  &      0      & 0       & 0    & 0 \\
       C_{QN}       &    0    &      0       &      I_{QQ} & 0       & 0    & 0 \\
       C_{RN}       &   0     &      0       &      0      & I_{RR}  & 0    & 0  \\
        v^T_{N}     &   0     &      0       &      0     & 0        & 1    & 0   \\
       w^T_{N}      &   0     &      0       &      0      & 0       & 0    & 1   
           \end{array}
    \right )
    \left (
           \begin{array}{c}
             x_N\\
             y_K\\
             y_L\\
             y_Q\\
             y_R \\
             y_v \\
             y_w
           \end{array}
    \right ) +
    \left (
           \begin{array}{c}
             e_N\\
             -e_K\\
             2e_L\\
             e_Q\\
             0  \\
             0  \\
             0 
           \end{array}
    \right )        \nonumber   \\             
                  \geq 0 \label{eqn4.14} \\
\end{equation}
\begin{equation}
\left (
           \begin{array}{ccccccc}
             x_N^T, & y_K^T, & y_L^T, & y_Q^T, & y_R^T & y_v & y_w 
           \end{array}            
    \right )  \geq 0                    \label{eqn4.15}
\end{equation} 

\begin{theorem}
\label{theorem4.3}  The following statements provide correct implications so the results are equivalent:
\begin{itemize}
\item[(a)]  The LCP (\ref{eqn4.123})-(\ref{eqn4.125}) has a  complementary solution,
\item[(b)]  The solution  determined is a solution to the satisfiablility problem or the solution found indicates that the satisfiability problem has no solution, so it is falsifiable, see definition (\ref{def2.5}).
\item[(c)]  The LP (\ref{eqn4.13})-(\ref{eqn4.15}) has an optimal solution $ (x^T) \geq 0 \quad 
\mbox{s.t.} \quad x_j = \{0,1 \}  \quad j = 1, 2, \cdots, n $ with $ y^T_{K}= y^T_{L}= y^T_{Q}= y^T_{R}=y_v = y_w = 0 $ so $ \hat{x_i} = \{0,1 \} \quad i= 1,2,\cdots, n $ is a solution to the Satifiability problem or the optimal solution determined for the LP (\ref{eqn4.13})-(\ref{eqn4.15}) contains one or more positive artificial variables $ y_J \geq 0, \not= 0 , \mbox{ for some}  \quad J \in \{K, L, Q, R\} $ and then the problem is falsifiable.
\end{itemize}
\end{theorem}

{\noindent \sc Proof}: [$(a) \rightarrow (b)$] The LCP (\ref{eqn4.123})-(\ref{eqn4.125}) satisfies the conditions of theorem \ref{theorem3.1}. If  the solution is such that $ \hat{x_i} = \{0,1 \} \quad i= 1,2,\cdots, N $ then by lemma  \ref{lemma2.1} the problem is satisfiable. On the other hand if one or more positive artificial variables $ y_J \geq 0, \not= 0 , \mbox{ for some} \quad   J \in \{K, L, Q, R\} $ then the LCP solution will not be a solution to the satisfiability problem, since the artificial variables will be basic in the solution.

[$(b) \rightarrow (c)$]  The LCP (\ref{eqn4.123})-(\ref{eqn4.125})  has a complementarity solution then by theorem \ref{theorem3.1} the LP (\ref{eqn4.13})-(\ref{eqn4.15}) has an optimal solution. Consider the complementarity solution $ y^T = \left ( (Diag(\hat{x})e) ^T, 0^T, 0^T, 0^T, 0^T, 0, 0  \right )$  which is feasible for the given LP  (\ref{eqn4.13})-(\ref{eqn4.15}) so the value of the objective function, after appropriate manipulation and cancellation  of the parameters in equations (\ref{eqn3.14}) and (\ref{eqn4.13}) will be:

\begin{equation}
\label {eqn4.211}
W^1 = \left (-e^T_{N}x_N,s^T_Ky_K, s^T_Ly_L, s^T_Qy_Q , s^T_Ry_R, v^T_Ny_v, w^T_Ny_w  \right ) < 0
\end{equation}

 and since  $ y_J = 0, \quad J \in \{K, L, Q, R\} $, the optimal solution by theorem \ref{theorem3.1},  so it is  a solution to the satisfiability problem. Other feasible solutions of the LP and complementatity solutions of the LCP can be derived but all will result in an increase in the objective function of the LP, as is immediate.

If the satisfiability problem is falsifiable then one or more artificial variables must enter the  basic solution and so an increase in one or more elements in some vector(s) $ y^T_K,  y^T_L,  y^T_Q,  y^T_R $ will occur, to render an optimal solution, but it cannot be a solution of the satisfiable problem since it is falsifiable.

[$(c) \rightarrow (a)$]  The  LP (\ref{eqn4.13})-(\ref{eqn4.15}) has  an optimal solution $ y^T = \left ( (Diag(\hat{x})e)^T, 0^T,0^T,0^T,0^T,0^T,0,0 \right ) $  and integer so  this is a solution to the Satisfiability problem, else the optimal solution  has one or more artificial variables in the optimal basis, so it cannot be a solution to the satisfiability probem but will be a falsifiable solution to the problem, see definition (\ref{def2.5}). So  Indeed the solution is the optimal solution to  LP (\ref{eqn4.13})-(\ref{eqn4.15}) by theorem \ref{theorem3.1} and will be a solution to the  satisfiability problem otherwise it will be a falsifiable solution. $\Box$

%% file: Complexity
\section{Complexity Results}    \label{a_Complexity}

The determination of the size of the problems to be encoded follow the definitions given in ~\cite{Schrijver86}.

Suppose that the original satisfiability problem is formed by $n$ literals and $ m$ clauses. This has to be transformed into a 3-sat formulation which depends on the number of literals per clause.  An upper bound to the number of clauses in the 3-sat formulation, given the original satisfiability problem, is $ nm $, while an upper bound to the number of variables $n + nm $ ~\cite{GM79}. 

Hence the size of the 3-sat problem to be solved is bounded by $ M $ clauses and $ N $ variables. 

Thus the LCP  (\ref{eqn4.123})-(\ref{eqn4.125})  based on the 3-sat problem formed from the original satisfiability problem will have size bounded by $M+N+2 $. Consequently, the  The LP (\ref{eqn4.13})-(\ref{eqn4.15}) to be solved will  consist of  $( M + N+2 )$ variables and inequalities. Moreover the data coefficients of the problem are integer or rational numbers.

Consider an interior point algorithm, to solve the  LP with a polynomial run time, even under degeneracy, to determine the optimal solution of an LP ~\cite{GC92} ~\cite{GO93} ~\cite{YY97}. 
 
\begin{theorem} 
\label{theorem3.4}
: The solution of any 3-satisfiability problem is a polynomial algorithm bounded by $O\left ((n + nm + 5)^{3.5}Size(C,b)\right )$  arithmetical operations, where n is the number of columns of the original satisfiability problem, which contains m clauses and $Size(C,B)$ is a polynomial of the number of bits required to encode the original satisfiability problem.
\end{theorem}
{\noindent \sc Proof}: A suitable interior point polynomial algorithm for solving the LP problem, requires  less than $O(p^{3.5}\sigma(B))$ operations, where p is the number of variables in the standard form of the problem and $\sigma(B)$ is the number of bits required to encode the LP problem.
We show that $p$ and $\sigma(B)$ are polynomially related to the size of the original problem. 
For the determinations of the size of each problem to be encoded, the size specification are applied \cite{Schrijver86}. 
 
Since C is a matrix with elements $ c_{ij} = -1, 0, +1 $, $ \quad  -1 \leq b_i \leq 2, \quad  \forall i,j = 1, 2, . . ., n $ yields the following bounds:
\begin{equation}
3m(n+1) \leq Size(C,b)\leq 3m(3n+1+\lceil \mbox{\rm \bf Log}_2(3)\rceil)    \label{eqn3.12bis}
\end{equation}
Approximate $ \lceil \mbox{\rm \bf Log}_2(a)\rceil$ as $ \lceil a \rceil $.
\begin{equation}
3m(n+1) \leq Size(C,b)\leq 3m(3n+4)    \label{eqn3.12tris}
\end{equation}
The size of the LP to be solved has $ ( nm+2)$ rows and $( n + nm+2)$ variables. Thus:
\begin{equation}
\mbox{Size ( M)} \leq ( n + nm+2)^2 \lceil {\bf Log}_2(3) \rceil \label{eqn3.53}
\end{equation}
The size of the vector $q$ of the affine term and $c$ the cost vector are, respectively:  
\begin{equation}
Size(q)= (  n + nm+2 +\lceil \mbox{\rm \bf Log}_2(3)\rceil)       \label{eqn3.54}
\end{equation}
\begin{equation}
\mbox{Size ( c)} \leq ((  n + nm+2 )(2 + \lceil \mbox{\rm \bf Log}_2(5(n+nm))\rceil)  \label{eqn3.55}
\end{equation}

Thus we obtain summing the the inequalities (\ref{eqn3.53}), (\ref{eqn3.54}),(\ref{eqn3.55}):
\begin{eqnarray}
 \mbox{ Size(M,q,c)} \leq 2(n +nm+5)^2  \nonumber \\   
\end{eqnarray}
 The ratio of the sizes is bounded by a linear function in the dimension of the problem, considering $ (n + nm + 1) \leq 3m(n + 1) $ is:
\begin{equation}
\frac{Size(M,q,c)}{Size(C,b)}\leq (n+nm+5)               \label{eqn3.13bis}
\end{equation}                                                                                    
Hence,  $\sigma(B)\leq ( n + nm + 5) Size (C,b)$ 
and so, the number of arithmetical operations required to solve the  LPs, will be at most, $O\left ((n + nm + 5)^{3.5}Size(C,b)\right )$ $\Box$

\begin{corollary} 
\label{corollary3.3}
 All propositional formulas in conjunctive normal form  belong to a language L in the class of polynomially bounded time complexity languages, {\bf P}.
\end{corollary}
Proof: All propositional formulas in conjunctive normal form can be recognized by an algorithm bounded by a polynomial number of arithmetic operations, by theorem \ref{theorem3.4}. Such an algorithm is equivalent to a deterministic Turing machine program  ~\cite{Khachian79} ~\cite{ Karmarkar84} ~\cite{GC92} ~\cite{GO93} ~\cite{YY97}.  By theorem  \ref{theorem4.3}, the program either finds a solution to the Satisfiability problem with an assignment of truth values or determines an optimal solution with a basic solution containing one or more positive elements of the vectors $ y^T_K,  y^T_L,  y^T_Q,  y^T_R $  indicating that no solution to the Satisfiability problem exists.$\Box$

%% file: Conclusions
\section{Conclusions}  \label{a_Conclusions}

The results proven in this paper show that the satisfiability problem is solved by an algorithm with a polynomially bounded number of arithmetical operations. 

The techniques used to prove this result are formal instrumentalist constructions in line with the concept of a formal language, rather than with realist motivations. It is felt that these formal deductive methods are important and useful to obtain general results, always a great concern of Science \cite{DJ77}.

%% file: ref.tex
\bibliography{biblio}